\documentclass[10pt,twoside,twocolumn]{article}
\usepackage{geometry}
\usepackage{amssymb,amsmath,amsfonts}
\usepackage{latexsym}
\usepackage{fancyhdr}
\usepackage{graphicx}
\usepackage{amsthm,multibox}
\usepackage{color}
\usepackage{hhline,xspace,rotate}
\usepackage{epsfig}
\usepackage{psfig}
\usepackage{pst-all}
\usepackage{pst-poly}
\usepackage{pst-coil}
\usepackage{multido}
\usepackage{url}
\usepackage{colortbl}
\usepackage{verbatim}

\usepackage{setspace}
\makeatletter

\usepackage[frenchb]{babel}

\renewenvironment{abstract}{\section*{Abstract}}{}

\begin{document}

\def\refname{References}
\def\figurename{{Figure}}
\def\tablename{{Table}}
\def\CaptionSeparator{\negthickspace:\space}
\title{{\bf Analysis of the $[{\bf 56},4^+]$ Baryon Masses in the $1/N_c$ Expansion\footnote{To appear in the proceedings of the JJC2004 --Journ\'ees Jeunes Chercheurs--, \^ile de Berder, France, 28 November--3 December 2004.}}}

\author{{\bf N. MATAGNE}\\
University of Li\`ege, Physics Department,\\
Institute of Physics, B.5, \\ 
Sart Tilman, B-4000 Li\`ege 1, Belgium\\
E-mail: nmatagne@ulg.ac.be}

\date{}

\maketitle

\newcommand{\fr}{\frac}
\def\1{\mbox{l\hspace{-0.53em}1}}
\begin{abstract}
\noindent We use the  $1/N_c$ expansion of QCD
to analyze the spectrum of positive parity resonances with strangeness
$S = 0, -1, -2$ and $-3$ in the 2--3 GeV mass region, supposed to belong 
to the $[\textbf{56},4^+]$ multiplet.  
We find that the spin-spin term brings the 
dominant contribution and that  the spin-orbit term is entirely negligible 
in the hyperfine interaction, in agreement with constituent quark model results. 
More data are strongly desirable, especially in the strange sector in order to 
fully exploit the power of this approach.

\end{abstract}

\section{Introduction}
\noindent Quantum chromodynamics (QCD), the theory of the strong interaction, is an  SU(3) gauge theory of quarks and gluons. 
Unfortunately, because of the complexity of the theory, it is not possible to solve it exactly.

\begin{figure}[h]
\begin{center}
\includegraphics*[width=7cm,keepaspectratio]{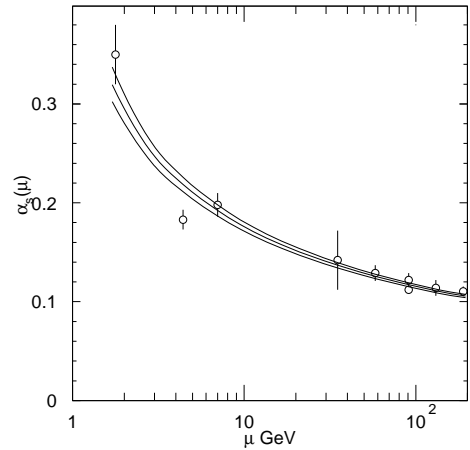}
\caption{Values of the strong coupling constant $\alpha_s(\mu)=g^2/4\pi$ vs. the energy scale $\mu$ \cite{PDG04}.}
\label{coupl}
\end{center}
\end{figure}

Generally, when a theory cannot be solved directly, one can try to make a perturbative expansion of the theory in terms of its coupling constant. However, the coupling constant $g$ of QCD becomes very large at low energies, too large for a perturbative treatement (see Figure \ref{coupl}). \\

In 1974, 't Hooft \cite{tHo74} suggested to generalize QCD from three to $N_c$ colors. He found that the inverse of the number of colors could be an expansion parameter of QCD.  
Later on, based on general arguments, Witten \cite{Wit79} analyzed the properties of mesons and baryon systems in this limit.

The $1/N_c$ expansion of QCD \cite{tHo74,Wit79} has been proved a useful approach 
to study baryon spectroscopy. It has been applied to the ground state
baryons \cite{DM93,DJM94,DJM95,CGO94,Jenk1,JL95,DDJM96} as well as to excited
states, in particular to the negative parity spin-flavor $[\textbf{70},1^-]$-plet 
($N = 1$ band)
\cite{Goi97,PY,CCGL,CGKM,CaCa98,SGS,Pirjol:2003ye}, to the positive parity Roper resonance
belonging to the $[\textbf{56'},0^+]$-plet  ($N = 2$ band) \cite{CC00} and to the  
$[\textbf{56},2^+]$-plet ($N = 2$ band) \cite{GSS03}. In this approach the main
features of the quark model emerge naturally and in addition new information
is provided as for example on the spin-orbit problem.

In this study we explore its applicability to the $[\textbf{56},4^+]$-plet 
($N = 4$ band) for the first time.
The number of experimentally known resonances in the 2--3 GeV region
\cite{PDG04}, expected to belong to this 
multiplet is quite restricted. Among the five possible candidates
there are two four-star resonances, $N(2220) 9/2^+$ and 
$\Delta(2420) 11/2^+$, one three-star resonance 
$\Lambda(2350) 9/2^+$, one two-star resonance 
$\Delta(2300) 9/2^+$ and one one-star resonance
$\Delta(2390) 7/2^+$.
This is an exploratory study 
which will allow us to make some predictions. 

The aim is to compute the mass operators for the $[\textbf{56}, 4^+]$ multiplet in a $1/N_c$ expansion.

This paper summarizes the results presented in Refs. \cite{Matagne:2004pm,MS2}.

\section{Baryons with $N_c=3$}
\noindent Baryons are color singlet bound states composed of three quarks\footnote{Here, we do not consider exotic baryons like pentaquarks which are composed of four quarks and one antiquark.}. 

One of the quantum numbers that characterises the quarks is the flavor. Six different flavors have been observed : $u$, $d$, $s$, $c$, $b$, $t$. Here, we consider only the three first ones.

The total wave function of baryons $\Psi$ is given by 
\begin{equation}
\Psi=\psi_{lm}\chi\phi C 
\end{equation}
where $\psi_{lm}$, $\chi$, $\phi$ and $C$ are the space, spin, flavor and color parts respectively. The color part $C$ is always antisymmetric, {\it i.e.}  a colorless state. So the remaining part must be symmetric because the total wave function must be antisymmetric as quarks are fermions.

One can introduce the SU$_f$(3) symmetry. Physically, if this symmetry is exact the mass of the three different quarks is the same, {\it i.e.} $m_u=m_d=m_s$.  

\begin{figure}[h]
\begin{center}
\includegraphics*[width=3cm,keepaspectratio]{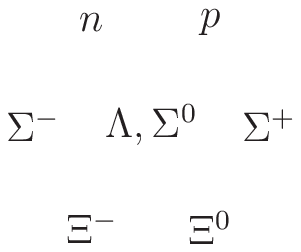}
\hspace{1cm}
\includegraphics*[width=3cm,keepaspectratio]{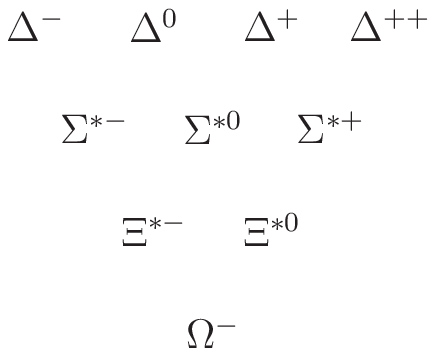}
\caption{SU$_f$(3) flavor diagrams for the octet on the left  and for the decuplet on the right.}
\label{flavor}
\end{center}
\end{figure}

One can classify baryons with respect to their flavor symmetry properties described by the irreducible representations of the SU$_f(3)$ group. We obtain the flavor diagrams of Figure \ref{flavor}. Particles belonging to the octet have a spin 1/2. For the decuplet, they have a spin 3/2. With exact SU$_f$(3), all the particles in each weight diagram have the same mass.

\section{SU(6) Symmetry}
\noindent One has shown that SU(6) is a good symmetry in $N_c \to \infty$ limit. That means that the 
first order operator of the expansion of the mass operator will be  SU(6) symmetric.

When we consider an exact SU(6) symmetry,  we assume that the mass of the particles that belong to the octet is the same that the ones that belong to the decuplet. In fact in SU(6) we have a spin symmetry and a flavor symmetry, {\it i.e.}
\begin{equation}
SU(6) \supset SU_S(2) \times SU_f(3).                                                                                 \end{equation}

The $SU(6)$ operators are:
\begin{eqnarray}
S^i & = & q^\dagger (S^i \otimes \1\, )q\nonumber \\
T^a & = & q^\dagger (\ \1 \otimes T^a) q \\
G^{ia} & = & q^\dagger (S^i \otimes T^a)q  \nonumber 
\label{su6gen}
\end{eqnarray}
where $S^i$ are the spin generators, $T^a$ the flavor generators, and $G^{ia}$ are the spin-flavor generators.

\section{Baryon Spectrum with a Linear Confinement and $N_c=3$}
\noindent It is possible to treat the $N_c=3$ baryon as a system of three non-relativistic quarks bound by a confining potentiel.
 
\begin{figure}
\begin{center}
\includegraphics*[width=5.5cm,keepaspectratio]{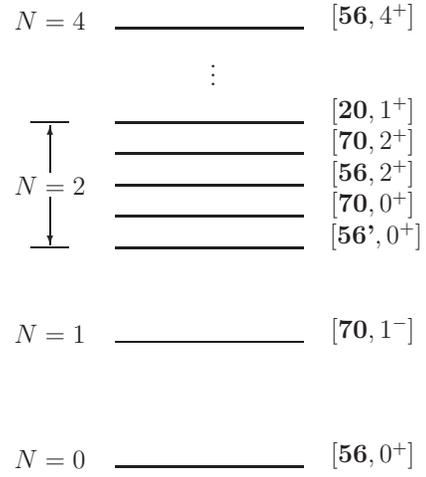}
\caption{The $N=0, 1$ and 2 levels of the baryon spectrum.}
\label{spectrum}
\end{center}
\end{figure}

In Figure \ref{spectrum}, we schematically draw the level sequence of the baryon spectrum up to the $N=2$ band with a linear confinement. To label the levels, the notation $[{\bf x},\ell^P]$ is used, where ${\bf x}$ represents the dimension of the SU(6) irreducible representations, $\ell$ the angular momentum of the states and $P$ the parity.

\section{The Mass Operator}
\noindent The mass spectrum is calculated in the $1/N_c$ expansion up to and including $\mathcal{O}(1/N_c)$ effects. The mass operator must be rotationally invariant, parity and time reversal even. We consider the SU$_f$(3) symmetry breaking to the first order. The isospin breaking is neglected. In term of the mass of the quarks, we assume that $m_u=m_d\not= m_s$.

From the $1/N_c$ expansion, one can write the mass operator as
\begin{equation}
M=\sum_{i=1}^3 c_iO_i + \sum_{i=1}^3 b_i \bar{B}_i+\mathcal{O}(1/N_c^2).
\label{mass}
\end{equation}
Here, $O_i$ ($i=1, 2, 3$) are SU$_f$(3) symmetry operators. The operators $\bar{B}_i$ ($i=1, 2, 3$) provide SU$_f$(3) breaking and are defined to have vanishing matrix elements for nonstrange baryons. The relation (\ref{mass}) contains the effective coefficients $c_i$ and $b_i$ as parameters. They represent reduced matrix elements that encode the QCD dynamics. The above operators and the values of the corresponding coefficients which we obtained from fitting the experimentally know masses are given in Table \ref{fit}.

\begin{table}
\renewcommand{\arraystretch}{1.25}
\begin{tabular}{llrrl}
\hline
\hline
Operator & \multicolumn{4}{c}{Fitted coef. (MeV)}\\
\hline
\hline
$O_1 = N_c \ \1 $                                               & \ \ \ $c_1 =  $  & 736 & $\pm$ & 30      $\ $ \\
$O_2 =\frac{1}{N_c} l_i  S_i$                                  & \ \ \ $c_2 =  $  &  4 & $\pm$ & 40   $\ $ \\
$O_3 = \frac{1}{N_c}S_i S_i$                                    & \ \ \ $c_3 =  $  &  135 & $\pm$ & 90   $\ $ \\
\hline
$\bar B_1 = -{\cal S} $                                         & \ \ \ $ b_1 = $  & 110 & $\pm$ & 67   $\ $ \\
$\bar B_2 = \frac{1}{N_c} l_i G_{i8}-\frac{1}{2 \sqrt{3}} O_2$  & \ \ \  \\

$\bar B_3 = \frac{1}{N_c} S_i G_{i8}-\frac{1}{2 \sqrt{3}} O_3$  & \ \ \ &  & &  \\

\hline \hline
\end{tabular}
\caption{Operators of Eq. (\ref{mass}) and coefficients resulting from the fit with $\chi^2_{\mathrm{dof}}\simeq 0.26$.}
\label{fit}
\end{table}

The matrix elements of $O_i$ and $\bar{B}_i$ ($i=1, 2, 3$) are given in Tables \ref{singlets} and \ref{octets}. The method followed to compute them is presented in Refs. \cite{Matagne:2004pm,MS2}.

\begin{table}[h!]
\[
\renewcommand{\arraystretch}{1.5}
\begin{array}{crrr}
\hline
\hline
           & \ \ \ \ \ \ \ \  O_1  & \ \ \ \ \ \ \ O_2  & \ \ \ \ \ \ \ O_3  \\
\hline
^28_{7/2}  & N_c  & - \fr{5}{2 N_c} & \fr{3}{4 N_c}  \\
^28_{9/2}  & N_c  &   \fr{2}{  N_c} & \fr{3}{4 N_c}  \\
^410_{5/2} & N_c  & - \fr{15}{2 N_c} & \fr{15}{4 N_c} \\
^410_{7/2} & N_c  & - \fr{4}{  N_c} & \fr{15}{4 N_c} \\
^410_{9/2} & N_c  &   \fr{1}{2 N_c} & \fr{15}{4 N_c} \\
^410_{11/2} & N_c  &   \fr{6}{  N_c} & \fr{15}{4 N_c} \\[0.5ex]
\hline
\hline
\end{array}
\]
\caption{Matrix elements of  SU$_f(3)$ singlet operators.}
\label{singlets}
\end{table}

\begin{table}
\[
\renewcommand{\arraystretch}{1.5}
\begin{array}{cccc}
\hline
\hline
  &    {\bar B}_1  &
      {\bar B}_2   &
      {\bar B}_3    \\
\hline
N_{J}       & 0 &                      0  &                           0  \\
\Lambda_{J} & 1 &   \fr{  \sqrt{3}\ a_J}{2 N_c} &   - \fr{3 \sqrt{3}}{8 N_c}   \\
\Sigma_{J}  & 1 & - \fr{  \sqrt{3}\ a_J}{6 N_c} &     \fr{  \sqrt{3}}{8 N_c}   \\
\Xi_{J}     & 2 &   \fr{ 2\sqrt{3}\ a_J}{3 N_c} &   - \fr{  \sqrt{3}}{2 N_c}   \\ [0.5ex]
\hline
\Delta_{J}  & 0 &                      0  &                           0  \\
\Sigma_{J}  & 1 &  \fr{ \sqrt{3}\ b_J}{2 N_c} &   - \fr{ 5 \sqrt{3}}{8 N_c}  \\
\Xi_{J}     & 2 &  \fr{ \sqrt{3}\ b_J}{  N_c} &   - \fr{ 5 \sqrt{3}}{4 N_c}  \\
\Omega_{J}  & 3 &  \fr{3\sqrt{3}\ b_J}{2 N_c} &   - \fr{15 \sqrt{3}}{8 N_c}  \\ [0.5ex]
\hline
\Sigma_{7/2}^8 - \Sigma^{10}_{7/2}    &  0 &  -\fr{  \sqrt{35}}{2 \sqrt{3} N_c}   &   0   \\
\Sigma_{9/2}^8 - \Sigma^{10}_{9/2}    &  0 &  -\fr{  \sqrt{11}}{  \sqrt{3} N_c}   &   0   \\
\Xi_{7/2}^8    - \Xi^{10}_{7/2}       &  0 &  -\fr{  \sqrt{35}}{2 \sqrt{3} N_c}   &   0   \\
\Xi_{9/2}^8    - \Xi^{10}_{9/2}       &  0 &  -\fr{  \sqrt{11}}{  \sqrt{3} N_c}   &   0   \\[0.5ex]
\hline
\hline
\end{array}
\]
\caption{Matrix elements of SU$_f(3)$ breaking operators. Here, $a_J = 5/2,-2$ for $J=7/2, 9/2$, respectively and
$b_J = 5/2, 4/3, -1/6, -2$ for $J=5/2, 7/2, 9/2, 11/2$, respectively.}
\label{octets}
\end{table}

\section{Fit and Discussion}

\noindent The fit of the  masses derived from  Eq. (\ref{mass})
and the available empirical values used in the fit,
together with the corresponding resonance status in the Particle Data Group
\cite{PDG04} are listed in Table \ref{multiplet}.

\begin{table*}[t]
\centering


\begin{tabular}{crrrrccl}\hline \hline
                    &      \multicolumn{4}{c}{Partial contribution (MeV)} &  Total (MeV)   & Empirical &  Name, status \\
\cline{2-5}
                    &      $c_1O_1$  &   $c_2O_2$ & $c_3O_3$ &  $b_1\bar B_1$   &   &  (MeV)   &    \\
\hline
$N_{7/2}$        & 2209 & -3 &  34 &   0  &   $ 2240\pm97 $  &   &  \\
$\Lambda_{7/2}$  &      &     &    & 110  &  $2350\pm118 $  & & \\
$\Sigma_{7/2}$   &      &     &    & 110  &  $2350\pm118 $  &               & \\
$\Xi_{7/2}$      &      &     &    & 220  &  $2460\pm 166$  &               &  \\
\hline
$N_{9/2}  $      & 2209 & 2   & 34 &   0  &   $2245\pm95 $  & $ 2245\pm65 $ & N(2220)**** \\
$\Lambda_{9/2}$  &  &     &    & 110  &  $ 2355\pm116 $  & $ 2355\pm15 $ &  $\Lambda$(2350)***\\
$\Sigma_{9/2}$   &      &     &    & 110  &  $ 2355\pm116 $  &               &                    \\
$\Xi_{9/2}$      &      &     &    & 220  &  $2465\pm164$  &               &                    \\
\hline
$\Delta_{5/2}$   & 2209 & -9  &168 &   0  &  $ 2368\pm175$  & &  \\
$\Sigma^{}_{5/2}$&      &     &    & 110  & $2478\pm187$  & &   \\
$\Xi^{}_{5/2}$   &      &     &    & 220  &  $2588\pm220$  &               &                   \\
$\Omega_{5/2}$   &      &     &    & 330  & $2698\pm266$  &               &                    \\
\hline
$\Delta_{7/2}$   &2209  &-5   &168 &  0   &  $2372\pm153$  & $2387\pm88$ &  $\Delta$(2390)* \\
$\Sigma'_{7/2}$  &      &     &    & 110  & $2482\pm167$  &               &                  \\
$\Xi'_{7/2}$     &      &     &    & 220  & $2592\pm203$  &               &                    \\
$\Omega_{7/2}$   &      &     &    & 330  &  $2702\pm252$  &               &                    \\
\hline
$\Delta_{9/2}$   &2209  & 1   &168 &  0   &   $2378\pm144 $  & $2318\pm132  $ &  $\Delta$(2300)**\\
$\Sigma'_{9/2}$  &      &     &    & 110  &   $2488\pm159$  &               &                   \\
$\Xi'_{9/2}$     &      &     &    & 220  &  $2598\pm197$  &               &                    \\
$\Omega_{9/2}$   &      &     &    & 330  &  $2708\pm247$  &               &                    \\
\hline
$\Delta_{11/2}$  &2209  &7    &168 &  0   &  $2385\pm164$  & $ 2400\pm100$ &   $\Delta$(2420)**** \\
$\Sigma^{}_{11/2}$ &    &     &    & 110  & $2495\pm177$  &               &                     \\
$\Xi^{}_{11/2}$  &      &     &    & 220  &  $2605\pm212$  &               &                     \\
$\Omega_{11/2}$  &      &     &    & 330  &  $2715\pm260$  &               &                     \\
\hline
\hline
\end{tabular}
\caption{Masses (in MeV)  
predicted by the $1/N_c$ expansion as compared with the 
empirically known masses. The partial contribution of each operator is indicated for
all masses. Those partial contributions in blank are equal to the one 
above in the same column.}
\label{multiplet}

\end{table*}

The values of the coefficients $c_i$ and $b_1$ obtained from 
the fit are presented in Table \ref{fit}, as already mentioned.

Due to the lack of experimental data in the strange sector it was 
not possible to include all the operators $\bar{B}_i$ in the fit in order 
to obtain 
some reliable predictions.  As the breaking of SU$_f$(3) is dominated by 
$\bar{B}_1$ we included only this operator in  Eq. (\ref{mass})
and neglected the contribution of the operators $\bar{B}_2$ and $\bar{B}_3$.
At a later stage, when more data will hopefully be available, all analytical 
work performed here could be used to improve the fit. That is why
Table \ref{fit} contains results for  $c_i$ ($i$ = 1, 2 and 3) 
and $b_1$ only. The $\chi^2_{\mathrm{dof}}$ of the fit is 0.26, where 
the number of degrees of freedom (dof) is equal to one (five data and four 
coefficients).

The first column of Table  \ref{multiplet}
 contains the 56 states (each state having a $2 I + 1$ multiplicity
from assuming an exact SU(2)-isospin symmetry).
 The columns two to five show the 
partial contribution of each operator included in the fit, multiplied by
the corresponding coefficient $c_i$ or $b_1$.
The column six gives the total mass according to Eq. (\ref{mass}). 
The errors shown in the predictions result from the errors on the 
coefficients $c_i$ and $b_1$ given in Table \ref{fit}.
As there are only five experimental data available, nineteen of these 
masses are predictions. 

The main question is, of course, how reliable is this fit. The answer
can be summarized as follows:
\begin{itemize}
\item[\textbullet] The main part of the mass is provided by the spin-flavor singlet operator
$O_1$, which is $\mathcal{O}(N_c)$. 

\item[\textbullet] The spin-orbit contribution given by $c_2O_2$ is small. This fact 
reinforces the practice used in constituent quark models where the spin-orbit
contribution is usually neglected. 
  
\item[\textbullet] The breaking of the SU(6) symmetry keeping the flavor symmetry exact
is mainly due to the spin-spin operator $O_3$. This hyperfine interaction 
produces a splitting between octet and decuplet states of approximately 130 MeV.

\item[\textbullet] As it was not possible to include the contribution of $\bar{B}_2$ and 
$\bar{B}_3$ in our fit, a degeneracy appears between $\Lambda$ and $\Sigma$. 
\end{itemize}

\section{Conclusions}
\noindent In conclusion we have studied the spectrum of highly excited resonances in the 2--3 GeV
mass region by describing them as belonging to the $[\textbf{56},4^+]$ multiplet.
This is the first study of such excited states based on the $1/N_c$ 
expansion of QCD. A better description should include multiplet 
mixing, following the lines developed, for example, in Ref. \cite{Go04}. 
 
Better experimental values 
for highly excited non-strange baryons as well as more data 
for the $\Sigma^*$ and $\Xi^*$ baryons are needed in order to understand
the role of the operator $\bar{B}_2$  within a multiplet
and for the octet-decuplet mixing. With better data the analytic work 
performed here will help 
to make reliable predictions in the large $N_c$ limit formalism.

\begin{center}\textbf{\large Acknowledgments}\end{center}{\large \par}

\noindent I gratefully thank Fl. Stancu for a careful reading of this report, J. L. Goity, C. Schat, N. N. Scoccola and P. Stassart for illuminating discussions,
and the organizers for the time spent for organizing 
this meeting. This work is supported by the Institut Interuniversitaire 
des Sciences Nucl\'eaires (IISN).

\end{document}